\documentclass{svmult}
\usepackage{amssymb}
\usepackage{amsmath}
\usepackage[square]{natbib} 
\usepackage{makeidx}     
\usepackage{multicol}    
\def\a{{\"a}}
\def\C{\ensuremath{\mathbb{C}}}
\def\H{\ensuremath{\mathbb{H}}}
\def\K{\ensuremath{\mathbb{K}}}
\def\N{\ensuremath{\mathbb{N}}}
\def\P{\ensuremath{\mathbb{P}}}
\def\R{\ensuremath{\mathbb{R}}}
\def\S{\ensuremath{\mathbb{S}}}
\def\Z{\ensuremath{\mathbb{Z}}}
\def\CCC{\ensuremath{\mathcal{C}}}
\def\FFF{\ensuremath{\mathcal{F}}}
\def\GGG{\ensuremath{\mathcal{G}}}
\def\HHH{\ensuremath{\mathcal{H}}}
\def\MMM{\ensuremath{\mathcal{M}}}
\def\SSS{\ensuremath{\mathcal{S}}}
\def\TTT{\ensuremath{\mathcal{T}}}
\def\XXX{\ensuremath{\mathcal{X}}}
\def\fs{\mathfrak{s}}
\def\fu{\mathfrak{u}}
\def\Om{\ensuremath{\Omega}}
\def\up{\ensuremath{\upsilon}}
\def\upn{\ensuremath{\upsilon^0}}
\def\Htzd{\ensuremath{H^2(Y^\delta,\Z)}}
\def\Htzt{\ensuremath{H^2(T,\Z)}}
\def\Htz{\ensuremath{H^2(X,\Z)}}
\def\Htrd{\ensuremath{H^2(Y^\delta,\R)}}
\def\Htrt{\ensuremath{H^2(T,\R)}}
\def\Htr{\ensuremath{H^2(X,\R)}}
\def\Hezd{\ensuremath{H^{even}(Y^\delta,\Z)}}

\def\Hez{\ensuremath{H^{even}(X,\Z)}}
\def\Herd{\ensuremath{H^{even}(Y^\delta,\R)}}

\def\Her{\ensuremath{H^{even}(X,\R)}}
\def\la#1{\label{#1}}
\def\ol{\overline}
\def\wh{\widehat}
\def\wt{\widetilde}
\def\no#1{\ensuremath{\langle#1,#1\rangle}}
\def\O{\ensuremath{\mathrm{O}}}
\def\PSL{\ensuremath{\mathrm{PSL}}}
\def\SL{\ensuremath{\mathrm{SL}}}
\def\SO{\ensuremath{\mathrm{SO}}}
\def\NS{\ensuremath{\mathrm{NS}}}
\def\Pic{\ensuremath{\mathrm{Pic}}}
\def\diag{\ensuremath{\mathrm{diag}}}
\def\im{\ensuremath{\mathrm{Im}}}

\def\rk{\ensuremath{\mathrm{rk}}}
\def\re{\ensuremath{\mathrm{Re}}}
\def\Span{\ensuremath{\mathrm{span}}}
\newcommand\req[1]{(\ref{#1})}
\newcommand\mb[1]{\mbox{\rm{}#1}}
\newlength{\myskip}
\newcommand\epr{\vspace{\myskip}
\ifmmode\qed\else{\unskip\nobreak\hfil
\penalty50\hskip1em\null\nobreak\hfil\qed
\parfillskip=0pt\finalhyphendemerits=0\endgraf}\fi
\noindent}
\newcommand\inv[2][1]{ {\textstyle {#1\over #2}} }
\renewcommand{\thefootnote}{\fnsymbol{footnote}\ifnum\value{footnote}=9
\setcounter{footnote}{0}\fi}
\makeindex

\begin{document} 
\title*{On Superconformal Field Theories\\ 
Associated to Very Attractive quartics}
\author{Katrin Wendland}
\institute{University of Warwick, Gibbet Hill, Coventry CV4-7AL, England /
UNC Chapel Hill, CB\#3250 Phillips Hall, Chapel Hill, NC 27599-3250, USA
\texttt{wendland@maths.warwick.ac.uk}}
\maketitle
\begin{abstract}
We study $N=(4,4)$ superconformal field theories 
with left and right
central charge $c=6$ which allow geometric
interpretations on specific quartic hypersurfaces in $\C{\P}^3$. Namely, we
recall the proof that the Gepner model $(2)^4$ admits a geometric 
interpretation on the Fermat quartic and give an independent cross-check
of this result,  providing a link to the
``mirror moonshine phenomenon'' on $K3$. We clarify the r\^ole of Shioda-Inose
structures\index{Shioda-Inose structure} in our proof and thereby generalize it: We introduce
\textsc{very attractive quartics} and show how on each of them a 
superconformal field theory can be constructed explicitly.
\end{abstract}
\index{Superconformal field theory!$N=(4,4)$ superconformal}
\index{Very attractive quartic}\index{$K3$-surface!very attractive}
\index{Gepner model}
\index{Mirror symmetry}
\index{Mirror moonshine}
\setcounter{minitocdepth}{2}
\dominitoc
\section*{Introduction}\la{wendland:intro}
Since the discovery of mirror symmetry\index{Mirror symmetry} \cite{wendland:lvw89,wendland:cogp91,wendland:grpl90} the 
quintic hypersurface in $\C{\P}^4$ 
has presumably become the most prominent
Calabi-Yau manifold \index{Calabi-Yau manifold} in theoretical physics. Its 
two-dimensional relative, the Fermat quartic
in $\C{\P}^3$, \index{Fermat quartic in ${\mathbb C}{\mathbb P}^3$}
has somewhat
eluded such fame.  However,
the study of two-dimensional Calabi-Yau manifolds seems rather promising
from a conformal field theoretic point of view:
On the one hand, the moduli space $\MMM_{SCFT}$ of superconformal
field theories (SCFTs) \index{SCFT|see{Superconformal field theory}}
\index{Superconformal field theory!moduli space} associated to Calabi-Yau two-folds can be defined
and studied on the level of abstract SCFTs, providing a sound basis
for a mathematical analysis. On the other hand, the algebraic structure
of $\MMM_{SCFT}$ is known explicitly, and its very description 
allows to draw links between geometry and SCFTs. Finally, 
to date only theories in
fairly low dimensional subvarieties  of $\MMM_{SCFT}$ have been 
constructed explicitly,  rendering $\MMM_{SCFT}$ a non-trivial object
to study.

In this note we investigate
the Fermat quartic, \index{Fermat quartic in ${\mathbb C}{\mathbb P}^3$}
and more generally so-called \textsc{very attractive
quartics},\index{Very attractive quartic}\index{$K3$-surface!very attractive} 
from a SCFT point of view. The main ideas arise as applications
of number theory and geometry to the study of SCFTs.

We start by giving an overview on the structure of the moduli space 
$\MMM_{SCFT}$ and\index{Superconformal field theory!moduli space} its geometric predecessors. In particular, we discuss
\textsc{attractive} and \textsc{very attractive} surfaces
(Defs.~\ref{wendland:attr}, \ref{wendland:vatt}),
providing\index{Attractive $K3$-surface}\index{$K3$-surface!attractive}
a first link to arithmetic number theory. Along the way the ``standard
torus'' $T_0$ and the Fermat quartic  $X_0\subset\C{\P}^3$ 
\index{Fermat quartic in ${\mathbb C}{\mathbb P}^3$}
serve as our favorite examples. 
To discuss SCFTs associated to $X_0$, in Sect.~\ref{wendland:gepner} we recall our
proof \cite[Thm.2.13, Cor.3.6]{wendland:nawe00} that the Gepner model\index{Gepner model} $(2)^4$ admits
a geometric interpretation on $X_0$, i.e.\ with the same complex structure
as $X_0$. We explain how the so-called
\textsc{Shioda-Inose-structures}\index{Shioda-Inose structure} give a guideline for the proof and
how orbifold constructions\index{Orbifold} allow us to also determine the normalized
K\a hler class, the volume, and the $B$-field\index{B-field} of that geometric interpretation.
In Sect.~\ref{wendland:phases} we give an independent cross-check of these results
by a careful application of Witten's analysis of phases in supersymmetric
gauge theories \cite{wendland:wi93} to the $K3$-case. We find that the 
\textsc{Fricke modular group $\Gamma_0(2)_+$} makes a natural appearance,
providing a link to the arithmetic properties of the mirror map 
\cite{wendland:nasu95,wendland:liya94}.
Sect.~\ref{wendland:SI} is devoted to possible generalizations of our proof
\cite[Thm.2.13, Cor.3.6]{wendland:nawe00}. First, the r\^ole of 
Shioda-Inose structures\index{Shioda-Inose structure}
is clarified. Then, for every very attractive quartic $X$\index{Attractive $K3$-surface}\index{$K3$-surface!attractive} 
we find an
$N=(4,4)$ SCFT $\CCC_X$ with $c=6$ which admits a geometric interpretation on
$X$. In fact, $\CCC_X$ is a $\Z_4$ orbifold\index{Orbifold} of a toroidal SCFT. As opposed
to the known $\Z_2$ orbifold CFTs with geometric interpretation on $X$, we
conjecture that $\CCC_X$  has a geometric interpretation with
the full hyperk\a hler structure\index{Hyperk\"ahler structure}
given by the natural one 
on $X\subset\C{\P}^3$. We give evidence in favor of
our conjecture, if (as implied by \cite{wendland:wi93}) it holds for $(2)^4$.
We end with a discussion in Sect.~\ref{wendland:disc} and state
some open problems and implications, if in
Sect.~\ref{wendland:SI} we have indeed found ``very attractive SCFTs'', in general.
\subsubsection*{Acknowledgments}
It is a pleasure to thank Paul Aspinwall, Gavin Brown, Gregory Moore,
Werner Nahm, and Emanuel Scheidegger for helpful discussions. This note is
a clarification and extension of ideas contained in our paper \cite{wendland:nawe00}, 
and we wish to thank Werner Nahm for that collaboration.
\section{Moduli spaces associated to Calabi-Yau two-folds}\la{wendland:modsp}
\index{Calabi-Yau manifold!two-folds!moduli space} 
We are interested in unitary, two-dimensional
SCFTs with central charge $c=6$ which arise in string
theory. These theories are expected to have nonlinear sigma model 
realizations\index{Superconformal field theory!nonlinear sigma model}
on Calabi-Yau manifolds of complex dimension $2$, i.e.\ either on a 
complex two-torus $T=Y^{\delta=0}$ or on a 
$K3$-surface $X=Y^{\delta=16}$. In this note,
$T,\,X,\,Y^\delta$ ($\delta\in\{0,16\}$) will always denote the respective
diffeomorphism type of a real four-manifold, with all additional structure
to be chosen later. Recall the signature
$\tau$ and Euler characteristic $\chi$ of these surfaces,
$$
-\tau(Y^{\delta=0}=T)=0=\delta,\;\;
\chi(T)=0,  \quad
-\tau(Y^{\delta=16}=X)=16=\delta,\;\; 
\chi(X)=24,
$$
and the respective cohomology groups. They are equipped with the metric
$$
\alpha,\beta\in H^\ast(Y^\delta,\R)\colon\quad
\langle\alpha,\beta\rangle = \int_{Y^\delta} \alpha\wedge\beta,
$$
which is induced by the intersection form on homology under Poincar{\'e} duality:
$$
\delta\in\{0,16\}\colon\quad 
\left\{
\begin{array}{rcccccl}
\R^{3,3+\delta} &\cong& \Htrd &\supset& \Htzd &\cong& \Gamma^{3,3+\delta},
\\[5pt]
\R^{4,4+\delta} &\cong& \Herd &\supset& \Hezd &\cong& \Gamma^{4,4+\delta}.
\end{array}
\right.
$$
Here, $\Gamma^{p,q}$ denotes the standard even unimodular lattice of
signature $(p,q)$, and the choice of the above isomorphisms amounts to
the choice of a marking.
\addtocounter{example}{-1}
\begin{example}\la{wendland:topex}
On our \textsc{standard torus} $T_0:=\R^4/\Z^4$ we use real Cartesian 
coordinates $x_1,\dots,x_4$. A complex structure is introduced\index{Calabi-Yau manifold!two-folds!complex structure} by choosing
complex coordinates,
\begin{equation}\label{wendland:sttor}
T_0:=\R^4/\Z^4, \quad z_1:=x_1+ix_2,\;\;  z_2:=x_3+ix_4,\quad
z_k\sim z_k+1\sim z_k+i.
\end{equation}
Our \textsc{standard $K3$-surface} 
is the \textsc{Fermat quartic} 
\index{Fermat quartic in ${\mathbb C}{\mathbb P}^3$}
\begin{equation}\label{wendland:fermat}
X_0\colon\quad\quad z_0^4+z_1^4+z_2^4+z_3^4 = 0 \quad \mbox{in }\;
\C{\P}^3
\end{equation}
with the induced complex structure.\index{Calabi-Yau manifold!two-folds!complex structure}
\end{example}
The pre-Hilbert space of a SCFT associated\index{Superconformal field theory!$N=(4,4)$ superconformal}
to a Calabi-Yau two-fold provides a representation
of the $N=(4,4)$ superconformal algebra at $c=6$ which contains a left and
a right handed Kac-Moody algebra $\fs\fu(2)_l\oplus\ol{\fs\fu(2)}_r$,
such that all charges with respect to a Cartan subalgebra (i.e.\ all
doubled spins) are integral.
Although 
explicit constructions are known only for a small number of theories
associated to $K3$,
the  moduli space of such SCFTs\index{Superconformal field theory!moduli space}
has been determined to a high degree of
plausibility \cite{wendland:na86,wendland:se88,wendland:asmo94,wendland:nawe00}. It should be compared to the
known ``classical'' moduli spaces of  geometric structures\index{Calabi-Yau manifold!two-folds!moduli space} on 
Calabi-Yau two-folds. This section gives a summary of the
relevant results, most of which can be found in 
\cite{wendland:asmo94,wendland:as96,wendland:nawe00,wendland:we00,wendland:we01}. The mathematical background is  
beautifully explained in \cite{wendland:bpv84,wendland:as96}.
\subsection{Complex structures}\la{wendland:comp}
\index{Calabi-Yau manifold!two-folds!complex structure}
The choice of a complex structure on a Calabi-Yau two-fold $Y^\delta$ is
equivalent to the choice of a holomorphic volume form 
$\mu \in H^2(Y^\delta,\C)$ with $\mu \wedge\mu =0$, $\mu \wedge\ol\mu >0$.
The real and imaginary part $\Om_1,\,\Om_2\in\Htrd$ of $\mu $
hence span an oriented positive definite two-plane
$$
\Om:=\Span_\R\left\{ \Om_1,\,\Om_2\right\} 
\subset \Htrd \cong \R^{3,3+\delta}.
$$
In fact, by the Torelli theorem,\index{Torelli theorem} there is a $1\colon1$ correspondence
between such two-planes and points in the moduli space\index{Calabi-Yau manifold!two-folds!moduli space}
$\MMM^\delta_{cs}$ of complex structures\index{Calabi-Yau manifold!two-folds!complex structure} on $Y^\delta$. To describe
$\MMM^\delta_{cs}$, we fix a marking $\Htzd\cong\Gamma^{3,3+\delta}$
and express each two-plane $\Om$ in terms of $\Htzd$. This gives
a parametrization by the Grassmannian\footnote{For an inner product
space $W$ with signature $(p,q)$, ${\O}^+(W)$ denotes the component of ${\O}(W)$
which contains $\SO(p)\times {\O}(q)$. For $G\subset {\O}(W)$, 
$G^+:=G\cap{\O}^+(W)$.}
$$
\wt\MMM^\delta_{cs} = {\O}^+ \!\!\left( \Htrd \right) / \,
\left({\O}(2) \times {\O}(1,3+\delta)\right)^+.
$$
The dependence on the marking is eliminated by dividing out the appropriate
discrete group:
\begin{equation}\label{wendland:cs}
\MMM^\delta_{cs} = 
{\O}^+ \!\!\left( \Htzd \right)\backslash\,
{\O}^+ \!\!\left( \Htrd \right) /\, \, \left({\O}(2)\times {\O}(1,3+\delta)\right)^+.
\end{equation}
For the standard torus
$T_0$ as in \mb{\req{wendland:sttor}}, the holomorphic volume form is
$\mu _{T_0}=dz_1\wedge dz_2$, i.e.\
$$
\Om_{T_0} = \Span_\R \left\{ 
\Om_1=dx_1\wedge dx_3 + dx_4\wedge dx_2,\;
\Om_2=dx_1\wedge dx_4 + dx_2\wedge dx_3 \right\},
$$
such that $\Om_{T_0}$ is generated by lattice vectors 
$\Om_k\in H^2(T_0,\Z)$. Hence in this specific example,
the \textsc{N{\'e}ron-Severi group}
$\NS(T_0)=\Om_{T_0}^\perp\cap H^2(T_0,\Z)$, which for Calabi-Yau two-folds
agrees with the \textsc{Picard group}, has maximal rank 
$\rho(T_0)=6-2=4$. That is, $T_0$ is an \textsc{attractive}
abelian variety: \index{Attractive abelian variety}
\begin{definition}\la{wendland:attr}
Let $Y$ denote a Calabi-Yau two-fold with complex structure\index{Calabi-Yau manifold!two-folds!complex structure} given by
$\Om_Y\subset H^2(Y,\R)$ such that $Y$ has maximal \textsc{Picard number}
$\rho(Y):=\rk (\Pic(Y))$, $\Pic(Y)=\NS(Y)=\Om_Y^\perp\cap H^2(Y,\Z)$.
In other words, assume $\rho(Y)=\rk (H^2(Y,\Z)) - 2$. Then 
the complex surface $Y$ \mb{(}or its complex structure\index{Calabi-Yau manifold!two-folds!complex structure} $\Om_Y$\mb{)} is called an \textsc{attractive} surface\footnote{The mathematical
\index{Attractive surface} literature dubs such surfaces \textsc{singular}, which may easily
cause confusion. We therefore rather borrow the terminology
from \mb{\cite{wendland:mo98b,wendland:mo98a}}.}.
\end{definition}
The following results make these surfaces so attractive for us:
\begin{theorem}\la{wendland:quafo}\hspace*{\fill}\vspace*{-0.5em}
\begin{enumerate}
\item\mb{\cite{wendland:shmi74}}
To every attractive complex structure\index{Calabi-Yau manifold!two-folds!complex structure} $\Om_T\subset\Htrt$ on a real 
four-torus $T$ we associate the quadratic form $Q_T$
of the \textsc{transcendental
lattice} $\Om_T\cap \Htzt$. 
This gives a $1\colon1$ correspondence between
attractive complex two-tori and $\SL_2(\Z)$ equivalence classes of
positive definite even integral quadratic forms.
\item\mb{\cite{wendland:shin77}}
The same is true for attractive $K3$-surfaces.
\end{enumerate}
\end{theorem}
This means that in order to specify the complex structure of\index{Calabi-Yau manifold!two-folds!complex structure}
an attractive Calabi-Yau two-fold $Y$, 
it suffices to state the quadratic form $Q_Y$ of the 
transcendental lattice. Thm.~\ref{wendland:quafo} establishes a deep
connection between the geometry of Calabi-Yau two-folds and
the classification of positive definite even integral
quadratic forms, i.e.\ a connection between geometry
and number theory, which we shall make continuous use of in this work.
\begin{example}\la{wendland:stcomp}
For the standard torus $T_0$ with complex structure\index{Calabi-Yau manifold!two-folds!complex structure} \mb{\req{wendland:sttor}} one
checks $Q_{T_0} = \diag( 2,2 )$. For the Fermat quartic \mb{\req{wendland:fermat}}
we have\index{Fermat quartic in ${\mathbb C}{\mathbb P}^3$}
\end{example}
\index{Attractive $K3$-surface}
\index{$K3$-surface!attractive}
\index{$K3$-surface!very attractive}
\index{Very attractive $K3$-surface|see{Very attractive quartic}}
\begin{theorem}\la{wendland:attqu}\mb{\cite{wendland:in76}}
The Fermat quartic\index{Fermat quartic in ${\mathbb C}{\mathbb P}^3$} 
$X_0$ given by \mb{\req{wendland:fermat}} is attractive, with
quadratic form $Q_{X_0}=\diag(8,8)$ on the transcendental lattice.
\end{theorem}
In fact, Thm.~\ref{wendland:attqu}
shows that the Fermat quartic\index{Fermat quartic in ${\mathbb C}{\mathbb P}^3$} 
is \textsc{very attractive}:
\begin{definition}\la{wendland:vatt}
An attractive $K3$-surface $X$ is \textsc{very attractive} if the
associated quadratic form $Q_X$ obeys 
$$Q_X=\left(\begin{array}{cc} 8a&4b\\4b&8c \end{array}\right)$$ for
some $a,\,b,\,c\in\Z$.
\end{definition}
\subsection{Hyperk\a hler structures}\la{wendland:hypka}
\index{Hyperk\"ahler structure}

Calabi-Yau manifolds are K\a hler by definition. If 
$Y^\delta$ ($\delta\in\{0,16\}$) is equipped with a complex structure\index{Calabi-Yau manifold!two-folds!complex structure}
$\Om\subset\Htrd$, then by the Calabi-Yau theorem there is a 
$1\colon1$ correspondence between \textsc{K\a hler classes}
\index{K\"ahler class} \index{Calabi-Yau theorem}
$\omega\in\Om^\perp\cap\Htrd$, $\no{\omega}>0$, and K\a hler-Einstein
metrics on $Y^\delta$. 
The real Einstein metric
underlying a pair $(\Om,\omega)$ as above is specified by the positive
definite oriented three-plane $\Sigma:=\Span_\R(\Om,\omega)\subset\Htrd$,
up to the volume.
Since by considering $\Sigma$ instead of $(\Om,\omega)$ the quantity
$\no{\omega}$ becomes superfluous, 
we can use any positive multiple of $\omega$, and we call it a \textsc{normalized
K\a hler class}.\index{K\"ahler class}
Combining Calabi-Yau and Torelli theorem \index{Torelli theorem}
 one has a $1\colon1$ correspondence
between positive definite oriented 
three-planes $\Sigma\subset\Htrd$
and real Einstein metrics on $Y^\delta$
(including orbifold limits),\index{Orbifold} up to the volume. 
Given an Einstein metric $g$ on $Y^\delta$, the corresponding
three-plane $\Sigma_g$ specifies an ${\S}^2$ of compatible complex structures\index{Calabi-Yau manifold!two-folds!complex structure}
$\Om\subset\Sigma_g$, i.e.\ a unique hyperk\a hler structure.\index{Hyperk\"ahler structure}
 In fact, 
the associated Hodge star operator $\ast_g$ acts as
involution on $\Htrd$, and $\Sigma_g$ can be obtained as the 
$\ast_g$-invariant part of $\Htrd$. It should be kept in mind that to date 
there is no direct method
available which allows to reconstruct $g$ from $\Sigma_g$.

The moduli space $\MMM^\delta_{hk}$ of\index{Hyperk\"ahler structure} hyperk\a hler structures\index{Calabi-Yau manifold!two-folds!moduli space} on 
$Y^\delta$ is now obtained in complete analogy to the moduli space\index{Calabi-Yau manifold!two-folds!moduli space}
$\MMM^\delta_{cs}$ of complex structures,\index{Calabi-Yau manifold!two-folds!complex structure}
c.f.\ \req{wendland:cs}:
$$
\MMM^\delta_{hk} = 
{\O}^+ \!\!\left( \Htzd \right)\backslash\,
{\O}^+ \!\!\left( \Htrd \right) /\,\, \SO(3) \times {\O}(3+\delta).
$$
\begin{example}\la{wendland:sthk}
The hyperk\a hler structure\index{Hyperk\"ahler structure} for our standard torus \mb{\req{wendland:sttor}} is specified
by $\Sigma_{T_0}=\Span_\R(\Om_{T_0}, \omega_{T_0})$ with attractive $\Om_{T_0}$
as in Ex.~\mb{\ref{wendland:stcomp}}\index{Attractive abelian variety} and 
$$
i\left(dz_1\wedge d\ol z_1 + dz_2\wedge d\ol z_2 \right)
\quad\sim\quad  dx_1\wedge dx_2 + dx_3\wedge dx_4 \;=:\; \omega_{T_0}.
$$
Note that $\omega_{T_0}\in \Om_{T_0}^\perp\cap H^2(T_0,\Z)$ with
$\no{\omega_{T_0}}=2$.

Similarly, we specify the hyperk\a hler structure\index{Hyperk\"ahler structure} $\Sigma_{X_0}$
on the Fermat quartic\index{Fermat quartic in ${\mathbb C}{\mathbb P}^3$} \req{wendland:fermat}
by choosing as normalized K\a hler class\index{K\"ahler class} the class $\omega_{FS}$ induced by the Fubini-Study
metric on $\C{\P}^3$. 
Then $\Sigma_{X_0}:=\Span_\R(\Om_{X_0}, \omega_{FS})$
with attractive $\Om_{X_0}$ as in Thm.~\mb{\ref{wendland:attqu}}, and 
$\omega_{FS}\in \Om_{X_0}^\perp\cap H^2(X_0,\Z)$ with
$\no{\omega_{FS}}=4$.\index{Attractive $K3$-surface}\index{$K3$-surface!attractive}
\end{example}
\subsection{$N=(4,4)$ Superconformal field theories}\la{wendland:scft}
\index{Superconformal field theory!$N=(4,4)$ superconformal}
By the above, the specification of a hyperk\a hler structure\index{Hyperk\"ahler structure} on 
a Calabi-Yau two-fold $Y^\delta$ ($\delta\in\{0,16\}$)
by a three-plane $\Sigma\subset\Htrd$ is equivalent to the specification
of a real Einstein metric $g$ on $Y^\delta$, up to the volume, where $\Sigma$
is the $\ast_g$-invariant subspace of $\Htrd$. To incorporate the volume
$V\in\R^+$, one may note that $\ast_g$ also acts as involution on $\Herd$
and consider the $\ast_g$-invariant subspace, there. But not every 
positive definite oriented four-plane $x\subset\Herd$ can be interpreted
as $(+1)$-eigenspace of such a Hodge star operator. However, Aspinwall
and Morrison \cite{wendland:asmo94} noticed that after the choice of a grading for
$\Herd\cong\R^{4,4+\delta}$ into $H^{even}=H^0\oplus H^2\oplus H^4$ by
selecting $\Z$-generators $\upn,\,\up$ of $H^0(Y^\delta,\Z),\,
H^4(Y^\delta,\Z)$, there exists a natural projection from the Grassmannian
of all positive definite oriented four-planes in $\Herd$,
\begin{equation}\label{wendland:cover}
\wt\MMM^\delta_{SCFT} := 
{\O}^+ \!\!\left( \Herd \right) /\,\, \SO(4) \times {\O}(4+\delta),
\quad \delta\in\{0,16\},
\end{equation}
to the parameter space of Einstein metrics on $Y^\delta$:
With $\Htrd=(\upn)^\perp\cap\up^\perp\cap\Herd$,
\begin{equation}\label{wendland:deco}
\begin{array}{rcrcccl}
\wt\MMM^\delta_{SCFT} &\longrightarrow&
\wt\MMM^\delta_{hk}&\times &\R^+&\times &\Htrd,\\[5pt]
x&\longmapsto& \left(\Sigma\right.&,&V&,&\left.B\right),\\[5pt]
&&x&=&\multicolumn{3}{l}{\Span_\R\left\{ \xi(\Sigma), \;
\upn+B+\left(V-\inv{2}\no{B}\right)\up\right\},}\\[5pt]
&&&&&\multicolumn{2}{l}{\xi(\sigma) := \sigma-\langle\sigma,B\rangle\up
\;\mbox{ for }\; \sigma\in\Htrd.}
\end{array}
\end{equation}
Here, $\upn,\,\up$ can be characterized by
\begin{equation}\label{wendland:null}
\upn,\,\up\in\Hezd\colon\quad
\no{\upn}=\no{\up}=0,\;\; \langle\upn,\up\rangle=1.
\end{equation}
Note that this gives $\wt\MMM^\delta_{SCFT}$ the structure of a bundle
over the parameter space $\wt\MMM^\delta_{hk}\times \R^+$ of Einstein 
metrics on $Y^\delta$. We interpret $V\in\R^+$ as parameter of
volume, and the fiber coordinate $B\in\Htrd$ is known as 
the \textsc{$B$-field}.\index{B-field}
In fact, the right hand side of \req{wendland:deco} is the parameter space of 
nonlinear sigma models on $Y^\delta$. \index{Superconformal field theory!nonlinear sigma model} On the other hand, using the
representation theory of our $N=(4,4)$ superconformal algebra at $c=6$
and deformation theory of SCFTs,\index{Superconformal field theory!$N=(4,4)$ superconformal} one shows that each component of
the parameter space of
$N=(4,4)$ SCFTs with $c=6$ is isomorphic to a Grassmannian \req{wendland:cover},
see \cite{wendland:na86,wendland:se88,wendland:ce91,wendland:asmo94}. Finally, there is a natural warped
product metric on the right hand side of \req{wendland:deco} which allows to
interpret that map
as isometry between an irreducible component of
the parameter space of $N=(4,4)$ SCFTs at $c=6$,
equipped with the Zamolodchikov metric, \index{Superconformal field theory!Zamolodchikov metric}
and the parameter space of 
nonlinear sigma models on $Y^\delta$.\index{Superconformal field theory!nonlinear sigma model} Thus this interpretation is 
compatible with the deformation theory both of SCFTs on the left hand
side and of the geometric data on the right hand side of \req{wendland:deco}
\cite{wendland:asmo94}.

As for $\MMM_{cs}^\delta$ and
$\MMM^\delta_{hk}$, the moduli space\index{Calabi-Yau manifold!two-folds!moduli space} is obtained from its smooth universal
cover $\wt\MMM^\delta_{SCFT}$
by dividing out a
discrete group of lattice automorphisms \cite{wendland:na86,wendland:asmo94,wendland:nawe00}:
\begin{equation}\label{wendland:ms}
\MMM^\delta_{SCFT} = 
{\O}^+ \!\!\left( \Hezd \right)\backslash\,
{\O}^+ \!\!\left( \Herd \right) /\,\, \SO(4) \times {\O}(4+\delta).
\end{equation}
Summarizing, the moduli space  of $N=(4,4)$
SCFTs\index{Superconformal field theory!moduli space} at $c=6$ decomposes into two components\footnote{We can prove
that the component $\MMM_{SCFT}^0$ is unique, but we cannot exclude
the occurrence of multiple components $\MMM_{SCFT}^{16}$. However, 
to date no\index{Superconformal field theory!$N=(4,4)$ superconformal}
example of an $N=(4,4)$ SCFT with $c=6$ has been found to contradict
uniqueness of $\MMM_{SCFT}^{16}$,  and in the following we restrict attention 
to a single such component.} 
$\MMM^\delta_{SCFT}$, 
$\delta\in\{0,16\}$, and every $N=(4,4)$
SCFT with $c=6$ can be associated either to $K3$ or to the torus, depending
on the value of $\delta$. Indeed, given an $N=(4,4)$ SCFT $\CCC$ with $c=6$
one can determine $\delta$ by calculating the conformal field theoretic
elliptic genus of $\CCC$, \index{Elliptic genus} which agrees with the geometric elliptic genus
of $Y^\delta$.  The theories in 
$\wt\MMM^{tori}_{SCFT}=\wt\MMM^0_{SCFT}$ are called \textsc{toroidal 
SCFTs}.
Each component $\wt\MMM^\delta_{SCFT}$ of the parameter space
is a natural extension of the parameter space of Einstein
metrics on $Y^\delta$, such that every oriented positive definite 
four-plane $x\subset\Herd$ corresponds to a SCFT\footnote{By
abuse of notation we generally denote a four-plane in $\Herd$ and the
corresponding
$x\in{\O}^+\!\!\left( \Herd \right) /\, \SO(4) \times {\O}(4+\delta)$
by the same symbol.}. \index{Superconformal field theory!$N=(4,4)$ superconformal}
More precisely,  \req{wendland:ms} is
a partial completion of the actual moduli space
of SCFTs on $K3$, see \cite{wendland:wi95}:
\begin{eqnarray}\la{wendland:roots}
&&x\in {\O}^+ \!\!\left( \Her \right) /\,\, \SO(4) \times {\O}(4+\delta)
\;\;\mbox{ corresponds to a SCFT}\\
&&\quad\Longleftrightarrow
x\subset\Her \mbox{ s.th. } 
\left\{ e\in x^\perp\cap\Hez \bigm| \no{e}=-2\right\} = \emptyset.
\nonumber
\end{eqnarray}
Each
choice of generators $\upn,\,\up$ of $H^0(Y^\delta,\Z),\,
H^4(Y^\delta,\Z)$ with \req{wendland:null} specifies a \textsc{geometric interpretation}
$(\Sigma,V,B)$ as in \req{wendland:deco}. 
The discrete group 
${\O}^+ \!\!\left( \Hezd \right)$ acts transitively on all possible pairs
$(\upn,\up)$ with \req{wendland:null}. 
If a SCFT $\CCC$ in $\MMM^\delta_{SCFT}$
has geometric interpretation $(\Sigma,V,B)$ such that $\Omega\subset\Sigma$
gives the complex structure\index{Calabi-Yau manifold!two-folds!complex structure} of a surface $Y_0^\delta$ as in Ex.~\ref{wendland:stcomp},
then we say that $\CCC$ \textsc{admits a geometric interpretation on 
$Y_0^\delta$}. 
\begin{example}\la{wendland:stscft}
It is easy to determine the location of
every toroidal \index{Superconformal field theory!$N=(4,4)$ superconformal}
SCFT in $\MMM^{tori}_{SCFT}=\MMM^0_{SCFT}$ and to construct
all theories in $\MMM^{tori}_{SCFT}$ explicitly, 
see \mb{\cite{wendland:na86}}. E.g.,
the torus \mb{\req{wendland:sttor}} has volume $V_{T_0}=1$, and setting the $B$-field\index{B-field}
to zero, $B_{T_0}:=0$, we can construct a SCFT $\TTT_0$ corresponding to
\begin{equation}\label{wendland:sttocft}
\TTT_0\colon\quad\quad
x_{T_0} := \Span_\R\left(
\Om_{T_0},\; \omega_{T_0},\; \upn+\up
\right)\;\in\; \wt\MMM^{tori}_{SCFT},
\end{equation}
with $\Om_{T_0},\, \omega_{T_0}$ as in Exs.~\mb{\ref{wendland:stcomp}, \ref{wendland:sthk}}. 
On the other 
hand, it is not easy to determine the explicit location of an abstractly
defined SCFT\index{Superconformal field theory!$N=(4,4)$ superconformal} within $\MMM^{K3}_{SCFT}$. Similarly, only very few theories
in $\MMM^{K3}_{SCFT}$ have been constructed explicitly, so far. As to the
example of the Fermat quartic\index{Fermat quartic in ${\mathbb C}{\mathbb P}^3$} $X_0$, it has been conjectured
\mb{\cite{wendland:ge87,wendland:ge88}} that the Gepner model\index{Gepner model} $(2)^4$ should admit a geometric
interpretation on $X_0$. Much evidence in favor of this conjecture has
been collected, in particular \mb{\cite{wendland:wi93}}, and the following two sections
are devoted to an explanation of its proof.
\end{example}
\section{Geometric interpretation of the Gepner model $(2)^4$}\la{wendland:gepner}
\index{Gepner model}
In this section we give a summary of our proof \cite[Cor.3.6]{wendland:nawe00}
that the Gepner model $(2)^4$ admits a geometric interpretation on the Fermat\index{Fermat quartic in ${\mathbb C}{\mathbb P}^3$} 
quartic $X_0$.
In fact, in \cite{wendland:eoty89} it was conjectured that
$(2)^4$ agrees with
the $\Z_4$  orbifold of our standard toroidal\index{Orbifold}
theory $\TTT_0$, see \req{wendland:sttocft}, since the respective partition functions
agree. Orbifold constructions have been studied in detail
in \cite{wendland:nawe00,wendland:we00,wendland:we01}, and the interested reader is referred to these
papers for further explanations. Here, only the following result 
from \cite{wendland:nawe00,wendland:we00} is needed:
\begin{proposition}\la{wendland:ZMorb}
Let $M\in\{2,3,4,6\}$, and let $T=\R^4/\Lambda$ denote a real four-torus with 
Einstein metric $g$ which is $\Z_M$ symmetric. In other words, there are a 
complex structure\index{Calabi-Yau manifold!two-folds!complex structure} and
complex coordinates $(z_1,z_2)$ which are compatible with $g$ and 
such that $g$ is invariant under 
$$
\zeta_M\colon\quad (z_1,z_2)\longmapsto (e^{2\pi i/M}z_1, e^{-2\pi i/M}z_2),
\quad\zeta_M\Lambda=\Lambda,
$$
i.e.\ $\Sigma_g\subset\Htrt^{\Z_M}$. Let $V_T\in\R^+$ denote the volume of
$T$ and $B_T\in\Htrt^{\Z_M}$ a $B$-field,\index{B-field} specifying a toroidal SCFT\index{Superconformal field theory!$N=(4,4)$ superconformal}
$\TTT$. Then the $\Z_M$-action $\zeta_M$ 
extends to a symmetry of $\TTT$, and the
corresponding orbifold\index{Orbifold} CFT, denoted $\TTT/\Z_M$, has a
geometric interpretation $(\Sigma,V,B)$
on the $\Z_M$  orbifold limit $X=\wt{T/\Z_M}$ of $K3$, which is obtained
from $T/\Z_M$ by minimally resolving all orbifold singularities. 
Let $\pi\colon\, T\longrightarrow X=\wt{T/\Z_M}$ denote the rational map
obtained from the minimal resolution, then 
$$
(\Sigma,V,B) = \left(\pi_\ast\Sigma_g, \inv[V_T]{M}, 
\inv{M} \pi_\ast B_T+\inv{M}\check{B}_M\right),
$$
where
$\check{B}_M\in\Htz\cap \left(\pi_\ast \Htzt^{\Z_M}\right)^\perp$
is a fixed primitive lattice vector with $\no{\check{B}_M}=-2M^2$.
\end{proposition}
The following application of Prop.~\ref{wendland:ZMorb} is a helpful exercise,
see Sect.~\ref{wendland:SI}:
\begin{corollary}\la{wendland:Z4ms}
Let $a,\,b,\,c\in\Z$ such that 
$$Q_{a,b,c}:=\left(\begin{array}{cc} 8a&4b\\4b&8c \end{array}\right)$$ is
positive definite. Then there is a toroidal 
SCFT $\TTT_{a,b,c}$ with $\Z_4$  orbifold\index{Orbifold} $\TTT_{a,b,c}/\Z_4$
in $\MMM^{K3}_{SCFT}$ corresponding to a four-plane
$x_{a,b,c}\subset\Her$ such that the following
holds\footnote{To clear notations, we generally let $\oplus$ denote
the orthogonal direct sum.}: 
$x_{a,b,c}=\Om^\ast\oplus\mho_{a,b,c}$, where $\Om^\ast,\,\mho_{a,b,c}$ are 
two-planes in $\Her$,
such that $\Om^\ast\cap\Hez$, 
$\mho_{a,b,c}\cap\Hez$ have rank $2$ and quadratic forms
$Q_{\Om^\ast}=\diag(2,2)$ and $Q_{\mho_{a,b,c}}=Q_{a,b,c}$, respectively.
\end{corollary}
\begin{proof}
Consider $T=\R^4/\Lambda=E_1\times E_2$ with 
orthogonal real two-tori $E_k$ at radii $R_k$, $k\in\{1,2\}$. That is,
with respect to real Cartesian coordinates
$x_1,\dots,x_4$ and the basis $dx_1,\dots,dx_4$ of $H^1(T,\R)$,
the $\Z$-dual  $\Lambda^\ast\cong H^1(T,\Z)$ 
of $\Lambda$ is generated by $\mu _1,\dots,\mu _4$ such that
$$
\exists\,R_1,\,R_2\in\R^+\colon\quad\quad
\left( \mu _1,\dots,\mu _4\right)
= \diag(R_1,R_1,R_2,R_2)^{-1}.
$$
We choose a complex structure\index{Calabi-Yau manifold!two-folds!complex structure} by setting $z_1:=x_1+ix_2,\,
 z_2:=x_3+ix_4$, and a K\a hler class \index{K\"ahler class}
$\omega_{R_2^2/R_1^2}\sim i(dz_1\wedge d\ol z_1 + dz_2\wedge d\ol z_2)$:
$$
\begin{array}{rcrcrcl}
dx_1\wedge dx_3 + dx_4\wedge dx_2 
&\;\sim\;& 
\mu _1\wedge \mu _3 &+& \mu _4\wedge \mu _2 & =:& \Om_1,\\[4pt]
dx_1\wedge dx_4 + dx_2\wedge dx_3 
&\;\sim\;& 
\mu _1\wedge \mu _4 &+& \mu _2\wedge \mu _3 & =:& \Om_2,\\[4pt]
dx_1\wedge dx_2 + dx_3\wedge dx_4
&\;\sim\;& \mu _1\wedge \mu _2 &+& {R_2^2\over R_1^2}\mu _3\wedge \mu _4 & =:& 
\omega_{R_2^2/R_1^2}.
\end{array}
$$
Here, $\Lambda$ and
$\Sigma:=\Span_\R(\Om_1,\Om_2,\omega_{R_2^2/R_1^2})$ are invariant under 
$\zeta_4\colon\, (z_1,z_2)\longmapsto (iz_1, -iz_2)$, i.e.\ 
$\Sigma\subset\Htrt^{\Z_4}$, and $V_T=R_1^2R_2^2$ is the volume of $T$.
For $a,\,b,\,c\in\Z$ as in the claim, 
$a>0,\,4ac-b^2>0$, since $Q_{a,b,c}$ is positive definite.
Let 
$$
\inv[R_2^2]{R_1^2}:=a, \quad V_T=R_1^2R_2^2:=c-\inv[b^2]{4a},\quad
B_T:=-\inv[b]{2a}\omega_a.
$$
These data specify a toroidal SCFT\index{Superconformal field theory!$N=(4,4)$ superconformal} $\TTT_{a,b,c}$, and
in terms of \req{wendland:deco},
$$
x_T=\Span_\R\left\{ \Om_1,\,\Om_2,\,\omega_a-\langle\omega_a,B_T\rangle\up,\,
\upn+B_T+\left(V_T-\inv{2}\no{B_T}\right)\up\right\},
$$
where $\left(\langle\Om_i,\Om_j\rangle\right)_{i,j}=\diag(2,2)$,
$\no{\omega_a}=2a$. By Prop.~\ref{wendland:ZMorb}, $\TTT_{a,b,c}/\Z_4$ is given by 
$x=\Om^\ast\oplus\mho_{a,b,c}\subset\Her$, where
with respect to appropriate generators $\wh\up^0,\,\wh\up$ of $H^0(X,\Z),\,
H^4(X,\Z)$
\begin{eqnarray*}
\Om^\ast &=& \Span_\R\left\{ \pi_\ast\Om_1,\, \pi_\ast\Om_2\right\},\\
\mho_{a,b,c} &=& \Span_\R\left\{
\xi_1:=\wt\omega_a-\langle\wt\omega_a,\wt B\rangle\wh\up,\,
\xi_0:=\wh\up^0
+\wt B+\left(\inv[V_T]{4}-\inv{2}\no{\wt B}\right)\wh\up\right\}\\
&&\mb{with } \;
\wt\omega_a:=\pi_\ast\omega_a,\quad
\wt B := \inv{4} \pi_\ast B_T + \inv{4}\check B_4 
= -\inv[b]{8a}\wt\omega_a + \inv{4}\check B_4,
\end{eqnarray*}
and $\no{\check B_4}=-32$.
First, by the results of \cite{wendland:shin77}, $\Om^\ast=\Span_\R\left\{ \wt\Om_1,\, 
\wt\Om_2\right\}$ with $\wt\Om_k=\inv2\pi_\ast\Om_k\in\Htz$.
Since $\pi$ has degree $4$ by construction,
$\left(\langle\wt\Om_i,\wt\Om_j\rangle\right)_{i,j}=\diag(2,2)$ and
$\no{\wt\omega_a}=4\no{\omega_a}=8a$, such that
\begin{eqnarray*}
&&\mho_{a,b,c}\cap\Hez \\
&&\quad\quad
=\Span_\Z\left\{ \xi_1=\wt\omega_a+b\,\wh\up,\,
\xi_2:=4\xi_0+\inv[b]{2a}\xi_1=4\wh\up^0+\check B_4+(c+4)\wh\up\right\}
\end{eqnarray*}
with 
$\left(\langle\xi_i,\xi_j\rangle\right)_{i,j}=Q_{a,b,c}$. 
This proves the claim.
\epr\end{proof}
Instead of giving
a direct proof for the claim $(2)^4=\TTT_0/\Z_4$ of \cite{wendland:eoty89},
we use the following result as a helpful
guide, without applying it explicitly (see Sect.~\ref{wendland:SI}):
\begin{theorem}\la{wendland:shinstr}\mb{\cite{wendland:shin77}}
Let $X$ denote an attractive $K3$-surface, \index{Attractive $K3$-surface}
which by Thm.~\mb{\ref{wendland:quafo}} is 
specified by its associated quadratic form $Q_X$. Then $X$ allows a
symplectic $\Z_2$-action $\iota$ \mb{(}that is, $\iota$ induces a trivial
action on $\Om_X\subset\Htr$\mb{)}
such that $Y=\wt{X/\iota}$ is 
attractive with associated quadratic form $Q_Y=2Q_X$. Moreover,
$Y=\wt{X/\iota}$ gives  the \textsc{Kummer surface}\index{Kummer surface} $\wt{T/\Z_2}$
of the attractive two-torus $T$ \mb{(}see Prop.~\mb{\ref{wendland:ZMorb}}\mb{)} 
with associated\index{Attractive abelian variety}
quadratic form $Q_T=Q_X$.

The triple\footnote{Here and in the following, 
$\stackrel{n\colon1}{\longrightarrow}$ and
$\stackrel{n\colon1}{\longleftarrow}$ denote rational maps of degree $n$.}
$X\stackrel{2\colon1}{\longrightarrow}
Y=\wt{T/\Z_2}\stackrel{2\colon1}{\longleftarrow}T$
is called a \textsc{Shioda-Inose-structure} \index{Shioda-Inose structure} \mb{(}in short an 
\textsc{SI-structure}\mb{)}.
\end{theorem}
Now note that $X=\wt{T_0/\Z_4}$ is an attractive $K3$-surface, which by the
result of \cite{wendland:shin77} mentioned in the proof of Cor.~\ref{wendland:Z4ms} has 
associated quadratic form $Q_X=\diag(2,2)=Q_{T_0}$. 
Hence by Thm.~\ref{wendland:shinstr} there is an SI-structure 
$X=\wt{T_0/\Z_4}\stackrel{2\colon1}{\longrightarrow}
\wt{T_0/\Z_2}\stackrel{2\colon1}{\longleftarrow}T_0$, and 
$Y=\wt{T_0/\Z_2}$ can be obtained from 
$X=\wt{T_0/\Z_4}$ by a $\Z_2$  orbifold\index{Orbifold}
construction. If $(2)^4=\TTT_0/\Z_4$, and if all symplectic symmetries 
extend to symmetries of the corresponding SCFTs,\index{Superconformal field theory!$N=(4,4)$ superconformal} then we can expect
a $\Z_2$  orbifold of $(2)^4$ to agree with $\TTT_0/\Z_2$. Indeed:
\begin{proposition}\la{wendland:twohat}\mb{\cite[Thm.3.3]{wendland:nawe00}}
Let $\sigma:=[2,2,0,0]$ denote the Gepner phase symmetry which acts
as the parafermionic $\Z_2$ on each of the first two factors of $(2)^4$. 
Then $(\wh 2)^4:=(2)^4/\sigma$
agrees with $\TTT_0/\Z_2$.
\end{proposition}
It is now easy to find a $\Z_2$-action $\eta$ on $(\wh 2)^4$ such that
$(2)^4=(\wh 2)^4/\eta$. Moreover, the proof of Prop.~\ref{wendland:twohat} gives an 
explicit dictionary between CFT and geometric data of $(\wh 2)^4$
and $Y=\wt{T_0/\Z_2}$, respectively, which allows to show that $\eta$ is
induced by a symplectic
$\Z_2$-action $\eta$ on $Y$. One then checks that
$\wt{Y/\eta}=\wt{T_0/\Z_4}=X$, which together with Prop.~\ref{wendland:ZMorb}
proves
\begin{proposition}\la{wendland:gepZ4}\mb{\cite[Thm.3.5]{wendland:nawe00}}
The Gepner model\index{Gepner model} $(2)^4$ has a geometric interpretation 
$(\pi_\ast\Sigma_{T_0},V,B)$ on
$X=\wt{T_0/\Z_4}$ with volume $V=\inv{4}$ and $B$-field\index{B-field}
$B=\inv{4}\check B_4$, where 
$\check B_4\in\Htz\cap\left(\pi_\ast H^2(T_0,\Z)^{\Z_4}\right)^\perp$
and $\no{\check B_4}=-32$.
\end{proposition}
By Sect.~\ref{wendland:hypka}, we can now write the four-plane 
$x_{(2)^4}\in\wt\MMM^{K3}_{SCFT}$ that corresponds to $(2)^4$
in the form \req{wendland:deco}, using the data
found in Prop.~\ref{wendland:gepZ4}. Then we determine a new pair 
$(\upn_Q,\up_Q)$ of null vectors with \req{wendland:null} to rewrite $x_{(2)^4}$:
\begin{equation}\label{wendland:qudeco}
x_{(2)^4}=\Span_\R\left\{ \xi(\Sigma_Q), \,
\upn_Q+B_Q+\left(V_Q-\inv{2}\no{B_Q}\right)\up_Q\right\}.
\end{equation}
If there is a two-plane $\Om_Q\subset\Sigma_Q$ such that 
$\Om_Q\cap\Htz$ has rank $2$ and quadratic form $Q_{X_0}=\diag(8,8)$,
then Thms.~\ref{wendland:quafo}, \ref{wendland:attqu} show that $(2)^4$ admits a geometric
interpretation on the\index{Fermat quartic in ${\mathbb C}{\mathbb P}^3$} Fermat quartic. 
The explicit form of \req{wendland:qudeco}
moreover allows us to determine the normalized K\a hler class,\index{K\"ahler class} the
volume,  and the $B$-field\index{B-field} of this
geometric interpretation. Indeed, 
\begin{proposition}\la{wendland:gepqu}\mb{\cite[Thm.2.13, Cor.3.6]{wendland:nawe00}}
The Gepner model\index{Gepner model} $(2)^4$ admits a geometric interpretation on the
Fermat quartic\index{Fermat quartic in ${\mathbb C}{\mathbb P}^3$} $X_0\subset\C{\P}^3$ 
with normalized K\a hler class\index{K\"ahler class}
$\omega_Q\in H^2(X_0,\Z)$, 
$\no{\omega_Q}=4$, volume $V=\inv{2}$, and $B$-field\index{B-field}
$B=-\inv{2}\omega_Q$.
\end{proposition}
Note that $\omega_Q$ may agree with the natural K\a hler class\index{K\"ahler class} $\omega_{FS}$
on $X_0\subset\C{\P}^3$, $\no{\omega_{FS}}=4$,
but the above methods do not prove this.
\section{A cross-check from physics: Phases on $K3$}\la{wendland:phases}
Let us take a closer look at the origin of the conjecture that
$(2)^4$ should admit a geometric interpretation on\index{Fermat quartic in ${\mathbb C}{\mathbb P}^3$} 
the Fermat quartic.
We use Witten's analysis of linear sigma models \index{Superconformal field theory!linear sigma model}
\cite{wendland:wi93} and the
assumption that under the renormalization group flow
each such theory flows to a (maybe degenerate) SCFT\index{Superconformal field theory!$N=(4,4)$ superconformal}
at the infrared fixed point. This in particular implies that for 
every quartic hypersurface $X\subset\C{\P}^3$, there should exist
a family $\FFF_X=\left(\CCC_w, w=e^{r+ i\vartheta}\in\C\cup\{\infty\}\right)$
of $N=(4,4)$ SCFTs at $c=6$ with geometric interpretation on $X$, normalized
K\a hler class\index{K\"ahler class}  $\omega_{FS}$ induced by the Fubini-Study metric on 
$\C{\P}^3$, and $B$-field\index{B-field} $B=\beta\omega_{FS}$, $\beta\in\R$. Moreover, the theory
at $w=1$ violates \req{wendland:roots}, hence is not well-defined, whereas 
$w=\infty$ corresponds to a large volume limit of $X$. The family has maximal 
unipotent monodromy around $w=\infty$, monodromy of order $2$ around
$w=1$, and monodromy of order $4$ around $w=0$. In fact, for $X=X_0$, 
$w=0$ should give the Gepner model\index{Gepner model} $(2)^4$.

In the smooth universal covering space $\wt\MMM^{K3}_{SCFT}$ (see
\req{wendland:cover}), the lift of
each family $\FFF_{X}$ can be described within a fixed
geometric interpretation, i.e.\ we
once and for all choose null vectors
$\upn,\,\up$ that generate $H^0(X,\Z),\, H^4(X,\Z)$ to use \req{wendland:deco}. 
Above, we have also specified a complex structure\index{Calabi-Yau manifold!two-folds!complex structure} and a normalized
K\a hler class,\index{K\"ahler class} 
or equivalently the three-plane $\Sigma$. We focus on 
the Fermat quartic $X_0$ in\index{Fermat quartic in ${\mathbb C}{\mathbb P}^3$} the following, so 
$\Sigma_{X_0}=\Span_\R(\Om_{X_0},\omega_{FS})$ as in Ex.~\ref{wendland:sthk}.
Moreover, $B=\beta\omega_{FS}$,
which allows us to cast all four-planes in our family
into the following form:
\begin{eqnarray}\la{wendland:mho}
&&x=\Omega_X\oplus\mho_{\beta,V}\quad
\mbox{ where with }\quad \no{\omega_{FS}}=4,\;
\beta\in\R,\;V\in\R^+\colon\nonumber\\
&&\quad\quad\quad\mho_{\beta,V}=\Span_\R\{\omega_{FS} - 4\beta\up,\, 
\upn+\beta\omega_{FS}+(V-2\beta^2)\up\}.
\end{eqnarray}
The parameters $(\beta,V)\in\R\times \R^+$ of the theories
under investigation can be
conveniently combined into a parameter $\tau$ on the
upper half plane $\H$:
$$
\tau:= \beta+i\sqrt{\inv[V]{2}} \;\;\in\;\; 
\H=\left\{ z\in\C \bigm| \im(z)>0 \right\}.
$$
To reproduce Witten's ${\S}^2\simeq\C\cup\{\infty\}$ 
we now  have to divide out all dualities
that leave invariant the subvariety of $\wt{\mathcal M}^{K3}_{SCFT}$ 
given by the four-planes $x$ as in \req{wendland:mho}, and then compactify
in a natural way.
Recall \cite{wendland:asmo94,wendland:nawe00} that all dualities on $\wt{\mathcal M}^{K3}_{SCFT}$
are generated by the geometric symmetries, which identify equivalent 
Einstein metrics, the integral $B$-field\index{B-field} shifts $B\mapsto B+\lambda,\,
\lambda\in H^2(X,\Z)$, and the Nahm-Fourier-Mukai transform
$\upsilon\leftrightarrow\upsilon^0$. In our case the geometric 
symmetries are of no relevance, since we have fixed a specific Einstein metric.
The $B$-field\index{B-field} shifts are restricted to $B\mapsto B+n\,\omega_{FS}, \, n\in\Z$,
by our constraint $B\sim\omega_{FS}$. The shift $B\mapsto B+\omega_{FS}$
induces the 
action\footnote{We cannot resist to use the notation $T$ for the map
$\tau\mapsto\tau+1$ and trust that this will cause no confusion with 
the same notation $T$ for the diffeomorphism type of a real four-torus in 
the rest of this note.}
$$
T:\quad \tau\;\;\longmapsto\;\;\tau+1
$$
on the parameter space $\H$. Finally, to study the Nahm-Fourier-Mukai transform note 
that  with \index{Nahm-Fourier-Mukai transform}
$$
\beta^*:=-{\beta\over V+2\beta^2}\;,\quad
V^*:={V\over (V+2\beta^2)^2}\;, 
$$
one has
$$
\mho_{\beta,V}=\Span_\R\left\{
\omega_{FS} - 4 \beta^*\upsilon^0,\, 
\upsilon+\beta^*\omega_{FS}+(V^*-2(\beta^*)^2)\upsilon^0
\right\}.
$$
So under $\upsilon\leftrightarrow\upsilon^0$ the new parameters are
$(\beta^*,V^*)$. Hence the Nahm-Fourier-Mukai transform acts 
on the parameter space $\H$ as
$$
S_2:\quad \tau=\beta+i \sqrt{\inv[V]{2}} 
\;\;\longmapsto\;\; {-\beta+i \sqrt{V/2}\over V+2\beta^2} = -{1\over 2\tau}.
$$
Let 
$$
\Gamma_0(2)_\ast := \langle T,\; S_2\rangle,
$$
then Witten's ${\S}^2$ should be realized as a compactification of
$\H/\Gamma_0(2)_\ast$. It is not hard to show that the fundamental domain
of $\Gamma_0(2)_\ast$ is 
\begin{equation}\label{wendland:fudo}
D:=
\left\{ 
\tau\in\H\,\left|\,
2|\tau|^2\geq1, 
|\re(\tau)|\leq \inv{2} \right.
\right\}
\end{equation}
and that $\Gamma_0(2)_\ast$ is  the normalizer 
group $\Gamma_0(2)_+$ of $\Gamma_0(2)$ in 
$\PSL_2(\mathbb R)$. 
$T$ identifies the two boundaries of $D$ at 
$|\re(\tau)|= \inv{2}$, and $S_2$ identifies $\tau$
with $-\ol\tau$ on the boundary given by the half circle $2|\tau|^2=1$.

Our observation that $\Gamma_0(2)_+$ gives
the automorphism group of the lattice generated by two null vectors
$\upn,\,\up$ with $\langle\upn,\up\rangle=1$
and $\omega$ with 
$\langle\upsilon,\omega\rangle=\langle\upsilon^0,\omega\rangle=0$,
$\no{\omega}=4$ agrees with \cite[Th.7.1]{wendland:do96}: For $N\in\N$, 
$\H/\Gamma_0(N)_+$ is nothing but the moduli space of $M_N$ polarized
$K3$-surfaces, \index{$K3$-surface!lattice polarized} 
i.e. $K3$-surfaces $X$ which possess $\omega\in\Pic(X)$
with $\no{\omega}=2N$.

From \req{wendland:fudo} it follows that 
$\H/\Gamma_0(2)_+$ can be naturally compactified by adding 
$\tau=i\infty$, such that
$$
{\S}^2\;\;\simeq\;\; \ol{ \H/\Gamma_0(2)_+}\;.
$$
Thus there are three special points in our ${\S}^2$, given by
$\tau=i\infty,\,\tau={i\over\sqrt2}$, and $\tau=-\inv{2}+{i\over2}$.
The  monodromies are generated by $T,\, S_2$, and $TS_2$,
respectively. Hence $\tau=i\infty$ is the point of maximal unipotent
monodromy and should give the point $w=\infty$ in $\FFF_X$. Indeed, this is
the point we have added to $\H/\Gamma_0(2)_+$ to compactify, and it
corresponds to $V\rightarrow\infty$;
\req{wendland:mho} shows that our four-plane then
degenerates to a plane spanned by $\Omega,\, \omega_{FS}$, 
and the null vector
$\upsilon$. In particular, the plane is independent of the value of $\beta$, as expected.
At $\tau={i\over\sqrt2}$ 
the monodromy has order $2$, i.e.\ gives a Weyl reflection
on cohomology as expected around the point $w=1$
in $\FFF_X$: Note the root
$e:=\upsilon-\upsilon^0,\, e\perp x$, which indeed violates \req{wendland:roots}. Finally,
$\tau=-\inv{2}+{i\over2}$ must correspond to the Gepner point $w=0$. 
Indeed, the values for volume and $B$-field\index{B-field} in this point are
$V=\inv{2}$ and $B=-\inv{2}\omega_{FS}$, in agreement with our explicit
calculation leading to Prop.~\ref{wendland:gepqu}. 

We find a $1\colon1$ map from our moduli space to Witten's ${\S}^2$
which   maps the special points correctly by setting
$$
r+i\vartheta
:= 2\pi\left( \ln \left( 2V-1\right) +  i\beta\right),\quad
w:=e^{r+i\vartheta}.
$$
The volume enters logarithmically, as expected from the
description of the moduli space $\mathcal M^{K3}_{SCFT}$\index{Superconformal field theory!moduli space} \cite{wendland:asmo94,wendland:di99}.

The above analysis gives an independent method to predict the full
geometric interpretation of the Gepner model\index{Gepner model} $(2)^4$ on the Fermat 
quartic.\index{Fermat quartic in ${\mathbb C}{\mathbb P}^3$}
It does not use mirror symmetry.\index{Mirror symmetry} However, let us now discuss
\paragraph{Mirror Moonshine on $K3$} 
\index{Mirror moonshine}
There is an interesting connection to arithmetic number theory which
relates the above analysis to the results of
\cite{wendland:nasu95,wendland:liya94}, 
see also \cite{wendland:veyu00}. Our calculation is independent of
the specific complex structure $\Om_X$ 
as long as $X\subset\C{\P}^3$. Hence
our family $\FFF_X\simeq{\S}^2$ is the
complexified K\a hler moduli space of the family of all quartic hypersurfaces
in $\C{\P}^3$. By mirror symmetry\index{Mirror symmetry} it should agree with
the complex structure moduli space of the mirror family, i.e.\ the
family $\left(\GGG_z, z=(4\psi)^{-4}\in\C\cup\{\infty\}\right)$
of $\Z_4^2$ orbifolds\index{Orbifold} of the quartic hypersurfaces
\begin{equation}\label{wendland:mirror}
z_0^4+z_1^4+z_2^4+z_3^4-4\psi z_0 z_1 z_2 z_3=0
\quad\mbox{ in }\quad \C{\P}^3,
\quad \psi\in\C\cup\{\infty\},
\end{equation}
where $\Z_4^2$ is generated by
$$
(z_0,z_1,z_2,z_3)\longmapsto(iz_0,-iz_1,z_2,z_3);\quad
(z_0,z_1,z_2,z_3)\longmapsto(iz_0,z_1,-iz_2,z_3),
$$
and in $\GGG_z$, all quotient singularities coming from fixed points
of the $\Z_4^2$ action are minimally resolved.
Here, $z=(4\psi)^{-4}$ is a true parameter of $\GGG_z$ since 
$z_0\mapsto-iz_0$ identifies the surfaces
at $\psi$ and $i\psi$.

The quartic \req{wendland:mirror} with $\psi=1$ 
has $16$ nodes, i.e. is
a Kummer surface\index{Kummer surface} whose
singularities have not been blown up. Hence $\GGG_z$ has monodromy of
order $2$ around $z=1/256$, which therefore corresponds 
to the point $\tau=\inv[i]{\sqrt2}$ in $\FFF_X$. 
By \req{wendland:mho} this means that the $\Z_4^2$ orbifold\index{Orbifold} of this
Kummer quartic has transcendental lattice $\diag(4,2)$, which uniquely 
determines it by Thm.~\ref{wendland:quafo}. 
The $16$ nodes form a single orbit
under the $\Z_4^2$-action, and hence in the $\Z_4^2$  orbifold give the 
unique root $e=\up-\upn$ with $e\perp x$ at 
$\tau=\inv[i]{\sqrt2}$ ($w=1$) mentioned above. Similarly, around
$z=\infty$, the family $\GGG_z$ has monodromy of order $4$. Hence $z=\infty$
corresponds to the Gepner point $\tau=-\inv{2}+\inv[i]{2}$ in $\FFF_X$, and
we find that the mirror\index{Mirror symmetry} of $(\Sigma_{X_0},V={1\over2}, 
B=-{1\over2}\omega_{FS})$ has transcendental
lattice $\diag(2,2)$, see \req{wendland:twotwo}.

The explicit mirror map for the family of quartics in $\C{\P}^3$ has 
been calculated in \cite{wendland:nasu95}. There, the group $\Gamma_0(2)_+$
makes its appearance already, and the discussion of its fixed points on $\H$
precisely matches our observations above. In \cite{wendland:liya94} these results
are rediscovered and extended, and it is pointed out that with
$q(\tau):=e^{2\pi i\tau}$,  $z=z(q)$, the function
$\tau\mapsto1/z -96$  is the Hauptmodul of $\Gamma_0(2)_+$. 
Given the r\^ole that $\Gamma_0(2)_+$ played in the above analysis,
this of course is
hardly a surprise.

The occurrence of Hauptmoduln in the mirror map has been entirely demystified
by C.~Doran \cite{wendland:do00a,wendland:do00b}: By the Torelli theorem,\index{Torelli theorem} the period domain of a 
one-parameter family of rank $19$ lattice polarized $K3$-surfaces\index{$K3$-surface!lattice polarized} 
$p\colon\XXX\longrightarrow S$ lies on a non-degenerate quadric in 
$\C{\P}^2$. Therefore, generalizing the $j$-function, one can define
a functional invariant $\HHH_M\colon S\longrightarrow\K_M$,
where $\K_M=\Gamma_M\backslash D_M$ denotes the coarse moduli space
of $M$-polarized $K3$-surfaces,\index{$K3$-surface!lattice polarized}  $D_M$ its smooth universal covering space,
and $\Gamma_M$ an arithmetic group. $\HHH_M$ is the composition of
the period morphism $S\longrightarrow D_M$ and the arithmetic quotient
$D_M\longrightarrow\K_M$. Doran shows that the Picard-Fuchs equation
for the family $\XXX$ is the symmetric square of a second order 
homogeneous linear Fuchsian ordinary differential equation. 
This perhaps is related to the existence of SI-structures as in \cite{wendland:pe86},
see \cite[5.1]{wendland:veyu00}.
The truncated projective period map gives the projective period ratio
$z(q)$ of that square root equation. The latter is the uniformizing 
differential equation for $S$ iff $\HHH_M$ is branched at most over the
orbifold\index{Orbifold} divisor in $\Pic(\K_M)$. Now on the one hand, $z(q)$ gives the
mirror map for $\XXX$, and on the other hand, by uniformization of
orbifold Riemann surfaces, $z(q)$ will be an automorphic function for a
finite index subgroup of a Fuchsian group of the first kind
iff the base curve $S$ is
so uniformized.
This result even generalizes directly to $n$-parameter families of rank
$20-n$ lattice polarized $K3$-surfaces\index{$K3$-surface!lattice polarized}  \cite{wendland:do00b}.

Summarizing, the beautiful arithmetic properties of the mirror maps\index{Mirror symmetry} of 
families of $K3$-surfaces \cite{wendland:liya94}
are neither restricted to the genus zero case,
nor to the one-parameter case, and they seem to be independent of 
moonshine.\index{Mirror moonshine}
\section{An application of Shioda-Inose structures?}\la{wendland:SI}
\index{Shioda-Inose structure}
Recall the idea underlying our proof of Prop.~\ref{wendland:gepZ4}:
Since there is an SI-structure 
$X=\wt{T_0/\Z_4}\stackrel{2\colon1}{\longrightarrow}
\wt{T_0/\Z_2}\stackrel{2\colon1}{\longleftarrow}T_0$, we predicted the 
existence of a $\Z_2$-type orbifold\index{Orbifold} $(\wh 2)^4=(2)^4/\sigma$ which could
be identified with  $\TTT_0/\Z_2$. However, having
found $\sigma=[2,2,0,0],\,\eta$ with $(\wh 2)^4=(2)^4/\sigma$ and
$(\wh 2)/\eta=(2)^4$ does not mean that $\sigma$ is induced by the symplectic
orbifold $X=\wt{T_0/\Z_4}\stackrel{2\colon1}{\longrightarrow}
\wt{T_0/\Z_2}$ from the SI-structure. In fact, in this section we will
argue the contrary. 

Assume that an SI-structure $X\stackrel{\mod\iota}{\longrightarrow}
Y\stackrel{\mod\pi}{\longleftarrow}T$ as in Thm.~\ref{wendland:shinstr}
can be lifted to the level of SCFTs.\index{Superconformal field theory!$N=(4,4)$ superconformal} This  implies that $X$ and $T$
admit $\iota$ and $\pi$-invariant Einstein metrics specified
by $\Sigma_X\subset\Htr^\iota$ and $\Sigma_T\subset\Htrt^\pi$, respectively.
Moreover, $\iota_\ast\Sigma_X=\pi_\ast\Sigma_T\subset H^2(Y,\R)$. However,
by the analysis of the \textsc{Nikulin involution} \index{Nikulin involution} $\iota$ performed in
\cite{wendland:mo93}, $(\iota_\ast\Sigma_X)^\perp\cap H^2(Y,\Z)=N\oplus E_8(-1)$,
where the lattice $E_8(-1)$ has intersection form the negative 
of the Cartan matrix of $E_8$, and
$N$ is generated by eight pairwise orthogonal roots $e_1,\dots,e_8$
and $\inv{2}\sum_{i=1}^8e_i$. On the other hand, by classical results on
Kummer surfaces, \index{Kummer surface}
$(\pi_\ast\Sigma_T)^\perp\cap H^2(Y,\Z)=\Pi$ is the
\textsc{Kummer lattice}  \index{Kummer lattice} which is not isomorphic to $N\oplus E_8(-1)$. 
Hence $\iota_\ast\Sigma_X\neq\pi_\ast\Sigma_T$.

This observation can be confirmed by a direct calculation: In 
\cite{wendland:shin77}, the branch locus of $X\stackrel{\mod\iota}{\longrightarrow}Y$
is described explicitly in terms of the lattice 
$\pi_\ast\Htzt\oplus\Pi\subset H^2(Y,\Z)$. In particular, the roots 
$e_1,\dots,e_8\in N$ are determined. They are the irreducible components
in the exceptional divisor for the minimal resolution of all singularities
in $X/\iota$. Then the map $\wt\eta$ reversing the orbifold\index{Orbifold} by $\iota$
should 
act as $\wt\eta_{\mid N}=-\mbox{id}$, $\wt\eta_{\mid N^\perp}=\mbox{id}$.
Hence with the results of \cite{wendland:shin77}, $\wt\eta$ can be calculated 
explicitly.  One checks that this map only leaves a subspace  of
signature $(2,3)$ of $\pi_\ast\Htrt$ invariant,
such that no Einstein metric $\Sigma_T\subset\Htrt^\pi$ can be found
with $\pi_\ast\Sigma_T\subset H^2(Y,\R)^{\wt\eta}$. In other words,
$\wt\eta$ does not extend to a symmetry of the $\Z_2$  orbifold  
$\TTT/\Z_2$ of any toroidal theory $\TTT$.

Summarizing, the symmetry $\sigma=[2,2,0,0]$ of $(2)^4$ which was used
in Sect.~\ref{wendland:gepner} is not induced by the Nikulin involution\index{Nikulin involution} $\iota$
of the SI-structure $\wt{T_0/\Z_4}\stackrel{\mod\iota}{\longrightarrow}
\wt{T_0/\Z_2}\stackrel{\mod\pi}{\longleftarrow}T_0$. But
recall that in Prop.~\ref{wendland:gepqu}
we proved that $(2)^4$ also has a geometric interpretation
on the Fermat quartic $X_0$. \index{Fermat quartic in ${\mathbb C}{\mathbb P}^3$}
Gepner had conjectured this
\cite{wendland:ge87,wendland:ge88} on the basis of a comparison between symmetries of $(2)^4$ 
and symplectic automorphisms of the quartic $X_0$. Witten's analysis
\cite{wendland:wi93} extends these ideas and provides us with a cross-check for
our results in Sect.~\ref{wendland:phases}. Under these identifications, 
$\sigma=[2,2,0,0]$ corresponds to the symplectic automorphism
$$
\sigma\colon\quad
(z_0,z_1,z_2,z_3)\longmapsto(-z_0,-z_1,z_2,z_3)
\quad\mbox{ in }\quad \C{\P}^3.
$$
Hence $(\wh 2)^4=(2)^4/\sigma$ can be expected to admit a 
geometric interpretation on $\wt{X_0/\sigma}$. The following theorem
implies that we have already proved this:
\begin{theorem}\mb{\cite[Thms.1,2]{wendland:in76}}\la{wendland:quaku}
An attractive $K3$-surface\index{Attractive $K3$-surface}\index{$K3$-surface!attractive} 
$X$ with associated quadratic form $Q_X$
is biholomorphic to a quartic surface
$$
X(f_1,f_2)\colon\quad
f_1(z_0,z_1)+f_2(z_2,z_3)=0 \quad\mbox{ in }\quad \C{\P}^3
$$
iff $X$ is very attractive,\index{Very attractive quartic}\index{$K3$-surface!very attractive} i.e.\ by Def.~\mb{\ref{wendland:vatt}}
iff $Q_X=4Q_T$ for some even integral
positive definite quadratic form $Q_T$.
In that case, let $Y=\wt{X/\sigma}$. Then $Y$ is an attractive $K3$-surface
with associated quadratic form $Q_Y$ such that
$Q_X=2Q_Y$. Moreover, $Y$ is 
canonically biholomorphic to the Kummer surface\index{Kummer surface} $\wt{T/\Z_2}$ where
$T=E_1\times E_2$ with $E_k\colon\, z_2^2=f_k(z_0,z_1)$ in $\C{\P}^2_{(1,1,2)}$
is an attractive torus\index{Attractive abelian variety} with associated quadratic form $Q_T$.
\end{theorem}
Indeed, since $Q_{X_0}=4Q_{T_0}$,
Thm.~\ref{wendland:quaku} implies that $\wt{X_0/\sigma}\simeq\wt{T_0/\Z_2}$,
which by Prop.~\ref{wendland:twohat} extends to 
$(\wh 2)^4=(2)^4/\sigma=\TTT_0/\Z_2$ on the level of SCFTs. In other words,
our proof that $(2)^4=\TTT_0/\Z_4$, which relied on the chain of orbifolds\index{Orbifold}
$(2)^4\stackrel{\mod\sigma}{\longrightarrow}(\wh 2)^4
\stackrel{\mod\eta}{\longrightarrow}(2)^4$, geometrically translates into
$Y=\wt{X_0/\sigma}\simeq\wt{T_0/\Z_2}$, $\wt{Y/\eta}\simeq\wt{T_0/\Z_4}$.
On the level of attractive complex structures we observe that the 
associated quadratic forms transform as
\begin{equation}\label{wendland:chain}
Q_{X_0}=\diag(8,8)
\stackrel{\mod\sigma}{\longrightarrow}
Q_{\wt{T_0/\Z_2}}=\diag(4,4)
\stackrel{\mod\eta}{\longrightarrow}
Q_{\wt{T_0/\Z_4}}=\diag(2,2).
\end{equation}
Now recall from our discussion in Sect.~\ref{wendland:phases} that $(2)^4$
should be given
by a four-plane $x_{(2)^4}\in\MMM^{K3}_{SCFT}$ with 
$x_{(2)^4}=\Om_{X_0}\oplus\mho_{-{1\over2},{1\over2}}$, 
where $\Om_{X_0},\,\mho^\ast:=\mho_{-{1\over2},{1\over2}}$ 
are generated by lattice vectors
in $\Hez$ such that by \req{wendland:mho}
\begin{equation}\label{wendland:twotwo}
\mho^\ast=
\Span_\R\left\{ \omega_{FS}+2\up,\, \upn-\inv{2}\omega_{FS}\right\}
=\Span_\R\left\{ \omega_{FS}+\up-\upn,\, \up+\upn\right\}.
\end{equation}
Hence the associated quadratic forms are $Q_{{X_0}}=\diag(8,8)$ and
$Q_{\mho^\ast}=\diag(2,2)$, which are exchanged in our chain \req{wendland:chain}. 
Finally, observe that the four-plane 
$x_{(\wh2)^4}\in\MMM^{K3}_{SCFT}$ corresponding to $(\wh 2)^4=(2)^4/\sigma$
can be split into $x_{(\wh2)^4}=\wh\Om\oplus\wh\mho$, where 
$\wh\Om,\,\wh\mho$ are generated by lattice vectors
in $\Hez$ such that the associated quadratic forms are 
$Q_{\wh\Om}=\diag(4,4)$ and $Q_{\wh\mho}=\diag(4,4)$. Our 
identification $(\wh 2)^4=(2)^4/\sigma=\TTT_0/\Z_2$ is blind towards an 
exchange $\wh\Om\leftrightarrow\wh\mho$, so
our chain \req{wendland:chain} schematically corresponds to
$\Om_{X_0}\oplus\mho^\ast\rightarrow\wh\Om\oplus\wh\mho\sim\wh\mho\oplus\wh\Om
\rightarrow\mho^\ast\oplus\Om_{X_0}$. 

The above implies a possible generalization of our construction: 
By Thms.~\ref{wendland:quaku}, for $a,\,b,\,c\in\Z$ with $a>0$ and $4ac-b^2>0$, there
is a very attractive quartic\index{Very attractive quartic}\index{$K3$-surface!very attractive} $X_{a,b,c}\subset \C{\P}^3$ with 
quadratic form on the 
transcendental lattice given by 
\begin{equation}\label{wendland:abc}
Q_{a,b,c}=
\left(\begin{array}{cc} 8a&4b\\4b&8c \end{array}\right). 
\end{equation}
On the other hand,
our discussion in Sect.~\ref{wendland:phases} was independent of the specific complex
structure $\Om_{a,b,c}=\Om_{1,0,1}$ on the quartic hypersurface in $\C{\P}^3$
which we chose in order to investigate 
$(2)^4$. Hence for  the family
$\FFF_{X_{a,b,c}}\simeq\ol{ \H/\Gamma_0(2)_+}$
of $N=(4,4)$ SCFTs\index{Superconformal field theory!$N=(4,4)$ superconformal} on $X_{a,b,c}$, equipped with the normalized 
K\a hler class\index{K\"ahler class}  $\omega_{FS}$ of the Fubini-Study metric and with $B$-field\index{B-field} 
$B=\beta\omega_{FS}$, the  discussion in Sect.~\ref{wendland:phases}
shows that there is a point with 
monodromy of order $4$ at $\beta=-\inv{2},\,V=\inv{2}$. This point corresponds
to a SCFT with $x_{a,b,c}=\Om_{a,b,c}\oplus\mho^\ast\in\MMM^{K3}_{SCFT}$ 
where $\Om_{a,b,c},\,\mho^\ast$ are generated by lattice vectors such that
the associated quadratic forms are $Q_{a,b,c}$ and $Q_{\mho^\ast}=\diag(2,2)$.
Now recall from Cor.~\ref{wendland:Z4ms} that up to 
$\Om_{a,b,c}\leftrightarrow\mho^\ast$
these are the data that can be obtained
from a $\Z_4$ orbifold\index{Orbifold} construction.  
If $x_{a,b,c}$ gives the location of $\TTT_{a,b,c}/\Z_4$ in 
$\MMM^{K3}_{SCFT}$, a chain similar to \req{wendland:chain} should exist.
Indeed, 
\begin{proposition}\la{wendland:vattgeom}
Let $a,\,b,\,c\in\Z$ such that
$\TTT_{a,b,c}$ denotes the toroidal SCFT\index{Superconformal field theory!$N=(4,4)$ superconformal} specified in Cor.~\mb{\ref{wendland:Z4ms}}.
Namely, $\CCC_{a,b,c}=\TTT_{a,b,c}/\Z_4$ is given by 
$x_{a,b,c}=\Om^\ast\oplus\mho_{a,b,c}\in \MMM^{K3}_{SCFT}$, 
where $\Om^\ast\cap\Hez$ and
$\mho_{a,b,c}\cap\Hez$ have rank $2$ and associated quadratic 
forms $Q_{\Om^\ast}=\diag(2,2)$
and $Q_{\mho_{a,b,c}}=Q_{a,b,c}$ as in \mb{\req{wendland:abc}},
respectively. Then $\CCC_{a,b,c}$ admits a geometric interpretation on 
the very attractive $K3$-surface $X_{a,b,c}$ with\index{Very attractive quartic}\index{$K3$-surface!very attractive} 
associated quadratic form
$Q_{a,b,c}$, normalized K\a hler class\index{K\"ahler class} $\omega_Q\in H^2(X_{a,b,c},\Z)$ with
$\no{\omega_Q}=4$, volume $V=\inv{2}$, and $B$-field\index{B-field} $B=-\inv{2}\omega_Q$.
\end{proposition}
\begin{proof}
The proof is entirely analogous to the proof of Prop.~\ref{wendland:gepqu}.
We use the notations and results of the proof of Cor.~\ref{wendland:Z4ms}.
Hence $x_{a,b,c}=\Om^\ast\oplus\mho_{a,b,c}$ with
\begin{eqnarray}\label{wendland:quartint}
\Om^\ast\cap\Hez &=& \Span_\Z\left\{\wt\Om_1,\, \wt\Om_2\right\},\\
\mho_{a,b,c}\cap\Hez &=& \Span_\Z\left\{
\xi_1=\wt\omega_a+b\,\wh\up,\,
\xi_2=4\wh\up^0+\check B_4+(c+4)\wh\up\right\}.\nonumber
\end{eqnarray}
If 
$(\lambda_1,\dots,\lambda_4)$ denotes a 
$\Z$-basis of $\Lambda$ for 
$T=\R^4/\Lambda$, which is dual
to the basis $(\mu _1,\dots,\mu _4)$ of $\Lambda^\ast$
used in  the proof of Cor.~\ref{wendland:Z4ms},
then the $\Z_4$-action $\zeta_4$ on $T$
has fixed points at each $\sum_{k}\inv[i_k]{2}\lambda_k$ with $i_k\in\{0,1\}$.
By the results of \cite{wendland:nawe00}, there are 
$e_{(i_1,i_2,i_3,i_4)}\in\Htz\cap\left(\pi_\ast\Htz^{\Z_4}\right)^\perp$ 
with $\no{e_{(i_1,i_2,i_3,i_4)}}=-2$ 
and $\langle e_{(i_1,i_2,i_3,i_4)},\check B_4\rangle=2$
(Poincar{\'e} dual to classes in the exceptional divisor of the blow up of 
$\sum_{k}\inv[i_k]{2}\lambda_k$) such that 
$$
\begin{array}{rcl}
\upn_Q &:=&
\inv{2}\left(\wt\Omega_1 +\wt\Omega_2\right)
+ \inv{2}\left( e_{(0,1,0,1)}-e_{(0,1,1,0)}\right),\\[5pt]
\up_{Q} &:=&
\inv{2}\left(\wt\Omega_1 -\wt\Omega_2\right)
- \inv{2}\left( e_{(0,1,0,1)}-e_{(0,1,1,0)}\right)
\end{array}
$$
are lattice vectors.
One  checks that $(\upn_Q,\up_Q)$ obey \req{wendland:null}; in fact, 
$(\upn_Q,\up_Q)$ agree with the vectors used in the proof of Prop.~\ref{wendland:gepqu},
see \cite[(2.18)]{wendland:nawe00}.
To determine the decomposition \req{wendland:qudeco}, we observe that
$\xi(\Sigma_Q)=\up_Q^\perp\cap x_{a,b,c}$ is generated by 
$\xi(\omega_Q):=\wt\Omega_1+\wt\Omega_2,\,\xi_1,\,\xi_2$.
Moreover, $\xi_k\perp\upn_Q$ for
$k\in\{1,2\}$, and these vectors
generate a two-plane $\Om_Q\subset\Sigma_Q$ such that 
$\Om_Q\cap\Hez$ has quadratic form $Q_{a,b,c}$. 
By Thm.~\ref{wendland:quaku} this means that we obtain a geometric interpretation
of $\CCC_{a,b,c}=\TTT_{a,b,c}/\Z_4$ on $X_{a,b,c}$. Finally,
$$
\omega_Q=\xi(\omega_Q)-\langle\xi(\omega_Q),\upn_Q\rangle\up_Q
=2\wt\Om_2+e_{(0,1,0,1)}-e_{(0,1,1,0)}
\;\in\Hez
$$
obeys $\no{\omega_Q}=4$, and 
$\xi(\Sigma_Q)^\perp\cap x_{a,b,c}$ is generated by
$\xi_4=\inv{2}(\wt\Omega_1 -\wt\Omega_2)=\upn_Q-\inv{2}\omega_Q$,
as claimed.
\epr\end{proof}
By results of \cite{wendland:shin77} we know that each 
quartic is biholomorphic to a Kummer surface,\index{Kummer surface} if it is very attractive. 
\index{Very attractive quartic}\index{$K3$-surface!very attractive}
Since all orbifolds\index{Orbifold} of toroidal SCFTs\index{Superconformal field theory!$N=(4,4)$ superconformal} can be constructed explicitly,
this means that for every quartic $X\subset\C{\P}^3$ 
we can find a $\Z_2$  orbifold CFT $\TTT^\prime_X/\Z_2$ of a toroidal SCFT
which admits a 
geometric interpretation on $X$,
if $X$ is very attractive.
Prop.~\ref{wendland:vattgeom} states that we then can also find a 
$\Z_4$ orbifold CFT $\TTT_X/\Z_4$
which has geometric interpretation with the same complex structure, 
but in general will be different from $\TTT_X^\prime/\Z_2$: 
The normalized K\a hler classes, volumes, and $B$-fields\index{B-field} will disagree. 
Note, e.g., that  $(2)^4$ does not agree with
any $\Z_2$ orbifold\index{Orbifold} of a toroidal SCFT \cite[Sect.2.4]{wendland:nawe00}.
The arguments used 
in Sect.~\ref{wendland:phases} as well as the result of Prop.~\ref{wendland:vattgeom}
indicate that, as opposed to the $\Z_2$ orbifold model $\TTT_X^\prime/\Z_2$,
the $\Z_4$ orbifold $\TTT_X/\Z_4$  may give a model on 
the very attractive quartic $X$\index{Very attractive quartic}\index{$K3$-surface!very attractive} 
with the natural hyperk\a hler structure:\index{Hyperk\"ahler structure}
\begin{conjecture}\la{wendland:conj}
The two-plane $\Om^\ast$  in Prop.~\mb{\ref{wendland:vattgeom}} agrees with
\mb{\req{wendland:twotwo}}, i.e.\ we can find a geometric interpretation of
$\CCC_{a,b,c}$ on $X_{a,b,c}$ with normalized K\a hler class\index{K\"ahler class}
the class $\omega_{FS}$ of the
Fubini-Study metric in $\C{\P}^3$. In other words, we can construct an
$N=(4,4)$ SCFT\index{Superconformal field theory!$N=(4,4)$ superconformal} with $c=6$ on every very attractive quartic, equipped with 
its natural hyperk\a hler structure.\index{Hyperk\"ahler structure}
\end{conjecture}
To sustain Conj.~\ref{wendland:conj} recall that
the analysis of 
Sect.~\ref{wendland:phases} predicts the model at $w=0$ in $\FFF_{X_{a,b,c}}$ to have
parameters $V=\inv{2},\, B=-\inv{2}\omega_{FS}$. The monodromy around $w=0$ has
order $4$, which is consistent with the model having a 
$\Z_4$ orbifold\index{Orbifold} interpretation $\TTT_{a,b,c}/\Z_4$ as in Prop.~\ref{wendland:vattgeom}.
In fact, assume that $(2)^4$ has a geometric interpretation
$\left(\Sigma_{X_0},V={1\over2},B=\right.$
$\left.-{1\over2}\omega_{FS}\right)$ as implied
by \cite{wendland:wi93}. The 
Fermat quartic\index{Fermat quartic in ${\mathbb C}{\mathbb P}^3$} $X_0\subset\C{\P}^3$ possesses a group 
$G\cong(\Z_4^3\rtimes\SSS_4)/\Z_4$ of symplectic automorphisms with 
Mukai number $\mu (G)=5$ (see \cite{wendland:mu88,wendland:as95}), which also leaves 
$\omega_{FS}$ invariant. This means that 
$\dim\left(H^2(X_0,\R)^G\right)=3$ and hence
$\Sigma_{X_0}= H^2(X_0,\R)^G$, as argued in \cite{wendland:as95}. 
On the other hand, Thms.~\ref{wendland:quafo}, \ref{wendland:attqu} and Prop.~\ref{wendland:vattgeom} 
imply $x_{(2)^4}=\Om_{X_0}\oplus\mho$ with $\Om_{X_0}=\mho_{1,0,1}$,
hence $\mho=\Om^\ast$. Although
this does not mean that the geometric
interpretation of $(2)^4$ in Props.~\ref{wendland:gepqu}, \ref{wendland:vattgeom}
agrees with the ``true quartic'' one, i.e.\ that $\upn=\upn_Q,\,\up=\up_Q$ 
with $\upn,\,\up$ as in \req{wendland:twotwo}, our assumptions ensure that a 
``true quartic'' interpretation by $\upn,\,\up$ exists.
The analysis of \cite{wendland:wi93} also implies an identification of some of
the deformations of the SCFT\index{Superconformal field theory!$N=(4,4)$ superconformal} $(2)^4$ with polynomial
deformations of the Fermat quartic $X_0$\index{Fermat quartic in ${\mathbb C}{\mathbb P}^3$} 
as in \cite{wendland:ge87}. 
Namely, identifying $\left(1_{\,\pm 1\, 0}^{\,\pm 1\,0}\right)$ in the 
$(j+1)^{\mbox{\small{}st}}$ component of $(2)^4$ 
with  $\alpha z_j$, $\alpha\in\C$, 
$$
V_\pm ^{(1)}:=
\left(2_{\,\pm 2\, 0}^{\,\pm 2\,0}\right)\left(2_{\,\pm 2\, 0}^{\,\pm 2\,0}\right)
\left(0_{\,0\, 0}^{\,0\,0}\right)\left(0_{\,0\, 0}^{\,0\,0}\right),
\quad
V_\pm ^{(2)}:=
\left(0_{\,0\, 0}^{\,0\,0}\right)\left(0_{\,0\, 0}^{\,0\,0}\right)
\left(2_{\,\pm 2\, 0}^{\,\pm 2\,0}\right)\left(2_{\,\pm 2\, 0}^{\,\pm 2\,0}\right)
$$
correspond to  deformations of $X_0$ as in \req{wendland:fermat} 
by $\alpha_1z_0^2z_1^2,\,\alpha_2z_2^2z_3^2$ with $\alpha_k\in\C$.

The $V_\pm ^{(k)}$ are  those $(\inv2,\inv2)$-fields which $(2)^4$
shares with $(2)^2\otimes(2)^2$ and  one of its subtheories
$(2)^2$. Since $(2)^2$ is the toroidal SCFT\index{Superconformal field theory!$N=(4,4)$ superconformal}
on $\R^2/\Z^2$  with vanishing $B$-field,\index{B-field} we have
$(2)^2\otimes(2)^2=\TTT_0$, and Prop.~\ref{wendland:gepZ4} implies
$(2)^4=(2)^2\otimes(2)^2/\Z_4$. Hence $V_\pm ^{(k)}$ give those 
$\Z_4$-invariant deformations in $\TTT_0$ which come from deformations
of one of the subtheories $(2)^2$, i.e.\
deformations of the radii $R_k$ of $E_k$, $k\in\{1,2\}$, in Cor.~\ref{wendland:Z4ms},
and $B$-field\index{B-field} deformations in 
$\Om_{T_0}^\perp\cap H^2(T_0,\Z)^{\Z_4}$.
In view of the four-plane $x_{(2)^4}$ with notations as in \req{wendland:quartint},
the volume-deformation of $T_0$
corresponds to a deformation of $\xi_2$ by
$\delta V:=\wh\up$, and the
deformation of $R_2/R_1$ corresponds to a deformation of 
$\xi_1$ by $\delta\wt\omega:=\wt\omega_2-\wt\omega_1$. Similarly,
all $V_\pm ^{(k)}$ give deformations of $\xi_1,\,\xi_2$ by
$\delta V,\,\delta\wt\omega$, respectively.

To show that Conj.~\ref{wendland:conj} holds in general if the above assumptions
hold for $(2)^4$, first note that in Prop.~\ref{wendland:vattgeom} we use the same 
$\Om^\ast$ for all $a,b,c$. We can maintain the ``true quartic'' 
geometric interpretation with $\upn,\,\up$ as in \req{wendland:twotwo} for 
all $a,b,c$, if $\upn$ and $\up$ (with $\upn,\up\perp\mho_{1,0,1}$)
are orthogonal to all $\mho_{a,b,c}$, i.e.\ to $\wh\up=\delta V$ and 
$\delta\wt\omega$. But the latter follows from the fact that $\delta V$ and 
$\delta\wt\omega$ give polynomial deformations of $X_0$ as in \req{wendland:fermat}
by complex multiples of $z_0^2z_1^2,\,z_2^2z_3^2$, i.e.\ deformations
of the complex structure\index{Calabi-Yau manifold!two-folds!complex structure} in the quartic interpretation \req{wendland:twotwo}
of $(2)^4$. 
Note that these polynomial deformations are also compatible with the form
$X(f_1,f_2)\colon\;f_1(z_0,z_1)+f_2(z_2,z_3)=0$ in $\C{\P}^3$ of every 
very attractive quartic\index{Very attractive quartic}\index{$K3$-surface!very attractive} as in Thm.~\ref{wendland:quaku}.
\section{Discussion}\la{wendland:disc}
We hope to have convinced the reader that the investigation of SCFTs\index{Superconformal field theory!$N=(4,4)$ superconformal} 
associated to $K3$ remains 
an interesting and challenging enterprize. The emphasis
of this work lies on a utilization of results in number theory and
geometry, specifically the \textsc{Shioda-Inose-structures}\index{Shioda-Inose structure} 
\cite{wendland:shmi74,wendland:shin77,wendland:in76,wendland:mo93}. That these structures should be useful in
the context of string theory had already been noticed in \cite{wendland:mo98b,wendland:mo98a}.

Another deep and interesting connection between number theory and SCFTs on 
$K3$ is the so-called \textsc{mirror moonshine phenomenon}\index{Mirror moonshine}
\cite{wendland:liya94,wendland:veyu00,wendland:do00b,wendland:do00a}. By a careful application of Witten's analysis
of phases in supersymmetric gauge theories \cite{wendland:wi93} to the $K3$-case we 
confirm that the \textsc{Fricke modular group $\Gamma_0(2)_+$} makes a natural
appearance in the study of the Fermat quartic 
$X_0\subset\C{\P}^3$.\index{Fermat quartic in ${\mathbb C}{\mathbb P}^3$} In 
particular, this gives a simple independent method to predict $B$-field\index{B-field} and
volume of the quartic interpretation of the Gepner model\index{Gepner model} $(2)^4$. This method
is also easily applied to the Gepner model\index{Gepner model} $(4)^3$, which should admit a 
geometric interpretation on the Fermat hypersurface in $\C{\P}^3_{(1,1,1,3)}$.
It would be interesting to extend these ideas to more-parameter cases. Since
Doran's analysis of the mirror moonshine\index{Mirror moonshine} phenomenon allows such an extension,
we hope that this will be possible. 

Though the idea of inverting Shioda-Inose structures\index{Shioda-Inose structure} by appropriate orbifold
constructions\index{Orbifold} is the guiding principle for our proof 
\cite[Thm.2.13]{wendland:nawe00} that the Gepner model\index{Gepner model} $(2)^4$ agrees with a 
$\Z_4$ orbifold, $(2)^4=\TTT_0/\Z_4$, we have not 
succeeded to perform such an inversion in general. In fact, we have shown
that such an inversion is impossible, as long as one works with a fixed
geometric interpretation, that is a fixed choice of grading in $\Her$. 
Since Shioda-Inose structures exist as geometric constructions for
arbitrary attractive $K3$-surfaces,\index{Attractive $K3$-surface}\index{$K3$-surface!attractive} 
the original idea suggested that a change
of geometric interpretation should not be necessary. Whether a combination
of orbifold\index{Orbifold} constructions and T-dualities can be found which leads to an
inversion of Shioda-Inose structures, thereby giving SCFTs\index{Superconformal field theory!$N=(4,4)$ superconformal} associated to
arbitrary attractive $K3$-surfaces, is left as an open problem for 
future work.

However, we have succeeded to generalize our method of proof for 
$(2)^4=\TTT_0/\Z_4$ in a different direction. Namely, for every very
attractive quartic $X$\index{Very attractive quartic}\index{$K3$-surface!very attractive} 
we find a $\Z_4$ orbifold\index{Orbifold} CFT $\TTT_X/\Z_4$ which 
admits a geometric interpretation on $X$. We conjecture that it even allows 
a geometric interpretation that
carries the natural hyperk\a hler structure\index{Hyperk\"ahler structure} which is induced
by the Fubini-Study metric on $\C{\P}^3$. In fact, we argue that if this
conjecture is true for $(2)^4$ and the Fermat 
quartic,\index{Fermat quartic in ${\mathbb C}{\mathbb P}^3$} 
where it is
generally believed, it should hold for all very attractive quartics. 
As a next step, our methods should imply that $\Z_4$ orbifolds\index{Orbifold}
can be used to construct SCFTs\index{Superconformal field theory!$N=(4,4)$ superconformal} with geometric interpretation on non-attractive
quartics $X(f_1,f_2)$ of the form 
$f_1(z_0,z_1)+f_2(z_2,z_3)=0$ in $\C{\P}^3$. The details
are left for a future publication.
%
%
%
%
\def\polhk#1{\setbox0=\hbox{#1}{\ooalign{\hidewidth
  \lower1.5ex\hbox{`}\hidewidth\crcr\unhbox0}}}
\providecommand{\bysame}{\leavevmode\hbox to3em{\hrulefill}\thinspace}

\printindex
\end{document}